\documentclass[iop]{emulateapj}

\usepackage{appendix,natbib}
\usepackage{amsmath}
\usepackage{subfigure}
\usepackage[backref,colorlinks=true,citecolor=blue,linkcolor=blue,urlcolor=blue]{hyperref}
\usepackage[all]{hypcap}

\usepackage{amsmath}
\usepackage{appendix}
\usepackage[utf8]{inputenc}
\usepackage{listingsutf8}
\usepackage{subfigure}

\newcommand{\halpha}{H\ensuremath{\alpha}}
\newcommand{\um}{$\mu$m}
\def\msun{\,${M_\odot}$}

\newcommand{\oh}{$12+\log({\rm O/H})$}
\slugcomment{Draft: {\today}. Accepted to ApJL}

\begin{document}

\title{Dependence of the IRX-$\beta$ dust attenuation relation on metallicity and environment\altaffilmark{*}}

\author{\sc Irene Shivaei\altaffilmark{1,2}, Behnam Darvish\altaffilmark{3}, Zahra Sattari\altaffilmark{4}, Nima Chartab\altaffilmark{4}, Bahram Mobasher\altaffilmark{4}, Nick Scoville\altaffilmark{3}, George Rieke\altaffilmark{1}}
\altaffiltext{1}{Steward Observatory, University of Arizona, Tucson, AZ 85721, USA}
\altaffiltext{2}{NASA Hubble fellow; email: ishivaei@arizona.edu}
\altaffiltext{3}{Cahill Center for Astrophysics, California Institute of Technology, Pasadena, CA 91125, USA}
\altaffiltext{4}{Department of Physics and Astronomy, University of California at Riverside, Riverside, CA 92521, USA}
\altaffiltext{*}{Based on observations made with the W.M. Keck Observatory, which is operated as a scientific partnership among the California Institute of Technology, the University of California, and the National Aeronautics and Space Administration.}

\begin{abstract}
We use a sample of star-forming field and protocluster galaxies at $z=2.0-2.5$ with Keck/MOSFIRE $K$-band spectra, a wealth of rest-frame UV photometry, and Spitzer/MIPS and Herschel/PACS observations, to dissect the relation between the ratio of IR to UV luminosity (IRX) versus UV slope ($\beta$) as a function of gas-phase metallicity ({\oh}\,$\sim 8.2-8.7$). We find no significant dependence of the IRX-$\beta$ trend on environment. However, we find that at a given $\beta$, IRX is highly correlated with metallicity, and less correlated with mass, age, and sSFR. We conclude that, of the physical properties tested here, metallicity is the primary physical cause of the IRX-$\beta$ scatter, and the IRX correlation with mass is presumably due to the mass dependence on metallicity. Our results indicate that the UV attenuation curve steepens with decreasing metallicity, and spans the full range of slope possibilities from a shallow Calzetti-type curve for galaxies with the highest metallicity in our sample ({\oh}\,$\sim 8.6$) to a steep SMC-like curve for those with {\oh}\,$\sim 8.3$. Using a Calzetti (SMC) curve for the low (high) metallicity galaxies can lead to up to a factor of 3 overestimation (underestimation) of the UV attenuation and obscured SFR. We speculate that this change is due to different properties of dust grains present in the ISM of low- and high-metallicity galaxies. 

\end{abstract}

\maketitle

\section{Introduction}

Dust attenuation and emission significantly modify our views of galaxies across cosmic time.
The ratio of dust emission in infrared (L(IR)) to the observed UV emission (L(UV)), known as the infrared excess or IRX, is a measure of the UV dust attenuation.
Additionally, the intrinsic UV continuum slope at $\lambda\sim 1600-2600$\,{\AA}, $\beta$, is another measure of UV dust reddening, as the intrinsic slope is almost constant with age for massive stellar populations where the birth and death rates are in equilibrium \citep{leitherer99}. Therefore, by calibrating $\beta$ using IRX, $\beta$ becomes a powerful empirical diagnostic for recovering the total (attenuation-corrected) UV luminosity, using rest-frame UV observations alone \citep[][hereafter, MHC99]{meurer99}. Such calibrations are central for high-redshift studies, where only optical/near-IR (i.e., rest-UV) observations have sufficient sensitivity to reach large samples of $L_*$ and sub-$L_*$ galaxies.

The effectiveness of an IRX-$\beta$ relation depends on its validity for different types of galaxies across cosmic time. While the original MHC99 IRX-$\beta$ relation applies to majority of galaxies, large scatters around this relation have been observed. 
Theoretical studies have shown that the IRX-$\beta$ scatter may depend on the type of dust, gas metallicity, star-formation history (SFH), dust-star geometry, and stellar population age \citep{popping17, safarzadeh17,narayanan18b,schulz20}. Observations have shown that the IRX-$\beta$ relation varies with stellar mass \citep{bouwens16c,reddy18a,fudamoto19,bouwens20}, IR luminosity \citep{buat12,casey14}, age \citep{siana09,reddy10,shivaei15a}, redshift \citep{capak15,pannella15}, and intrinsic $\beta_0$ \citep[$\beta$ for a dust-free system;][]{boquien12,reddy18a,schulz20}. These variations are linked to the diversity of galaxies' attenuation curves, seen in previous studies \citep[e.g.,][]{kriek13,scoville15,battisti20}. The attenuation curve variations stem from two main sources: (a) different geometrical distributions of dust with respect to stars (and different dust optical depths), and (b) different dust grain properties, which affect the shape of the underlying extinction curve irrespective of the dust-star geometry \citep[see the review by][]{salim20}. 
While scenario (a) is extensively studied by comparing the attenuation curves of galaxies with different dust optical depths (references above), scenario (b) is less explored. Dust compositions are related to gas-phase element depletions and abundances \citep{jenkins09}. Following this relation, \citet[][hereafter, S20]{shivaei20} studied scenario (b) by deriving the attenuation curve of $z\sim 2$ galaxies in two different metallicity ({\oh}) bins, while the dust optical depth distributions were kept the same. They found a steep SMC-like curve for galaxies with $12+\log({\rm O/H})\lesssim 8.5$, and a shallower \citet[][hereafter, C00]{calzetti00}-like curve at $12+\log({\rm O/H})\gtrsim 8.5$.

The S20 results were based on rest-frame UV and optical data. In this work, we use Spitzer/MIPS and Herschel/PACS IR data as direct tracers of dust emission for a unique sample of field and protocluster galaxies with Keck/MOSFIRE observations. Our goals is to study the dependence of the IRX-$\beta$ relation at $z\sim 2$ on metallicity, mass, age, and environment.

\begin{figure*}[t]
	\centering
		\includegraphics[width=.8\textwidth,trim={.2cm 0 0 0},clip]{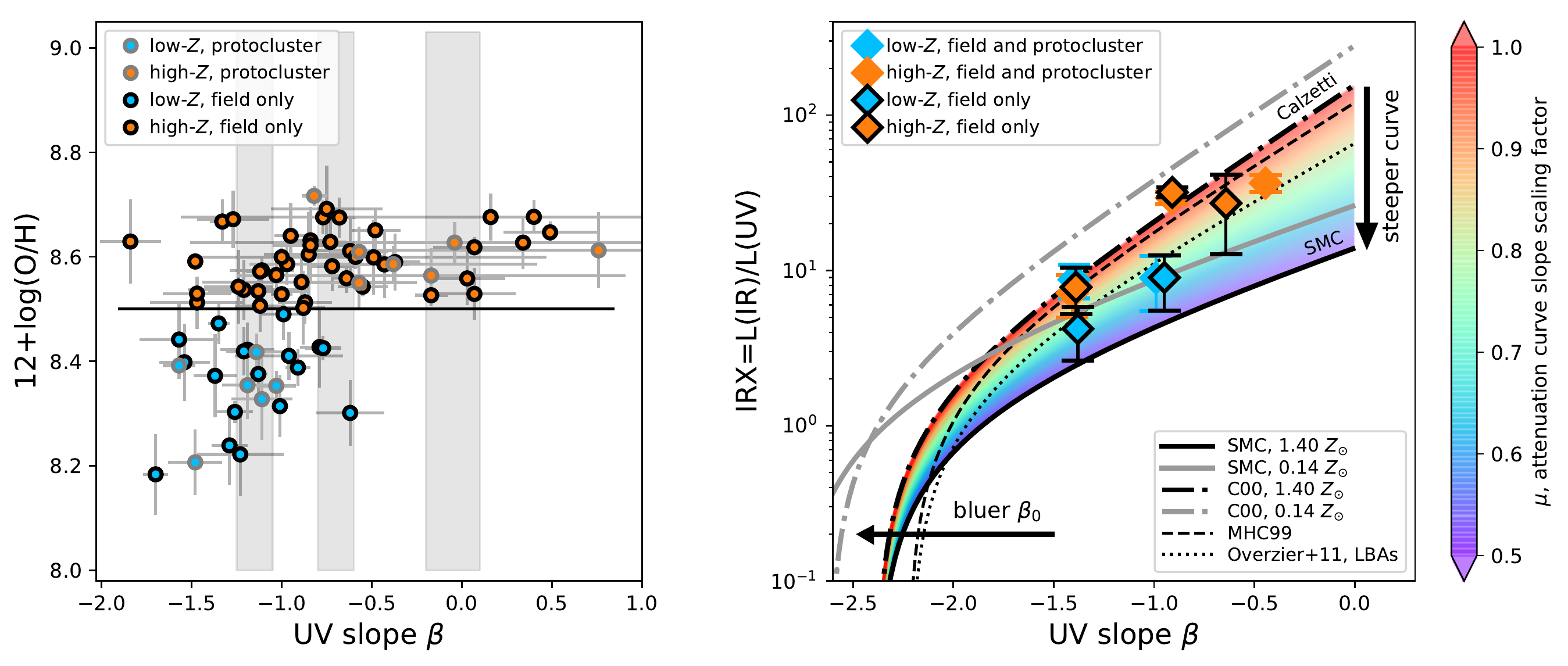} 
		\caption{{\bf Left:} Metallicity versus $\beta$ for the field and protocluster galaxies. The metallicity and $\beta$ boundaries used to divide the sample into the bins displayed in the right panel are shown with a black line and grey regions, respectively. $\beta$ boundaries are chosen randomly within the grey regions for 100 trials (see Appendix~\ref{ap:a3}). 
		{\bf Right:} IRX-$\beta$ for the full sample at $z=2.0-2.5$ (no edge-color symbols) and only field galaxies (black-edged symbols) at {\oh}\,$<8.5$ (low-$Z$, blue) and  {\oh}\,$>8.5$ (high-$Z$, orange). Galaxies with $\beta\gtrsim 0$ are not shown in the plot.	
		The original MHC99 curve \citep{meurer99} is shifted by 0.24\,dex upward \citep{reddy18a} to convert the far-IR luminosity in that study to the total IR luminosity assumed here.
		The SMC \citep{gordon03} and \citet[][C00]{calzetti00} relations are taken from \citet{reddy18a}, assuming a constant star formation history and an age of 100\,Myr for stellar populations with two different stellar metallicities (0.14 and 1.40\,$Z_{\odot}$). A lower stellar metallicity results in a bluer intrinsic $\beta_0$ and hence, shifts the IRX-$\beta$ curves to the left. At a fixed $\beta_0$, corresponding to that of the 1.40\,$Z_{\odot}$ model, we show the loci of attenuation curves with different slopes in between the C00 and SMC slopes, with rainbow colors. The colors show the power-law slope deviation factor, $\mu$, where $\mu=1$ corresponds to the C00 curve (black dash-dotted line) and the steepest curve shown has a $\mu=0.5$, resembling an SMC-like curve (black solid line). These curves are reproduced by scaling the exponent of the C00, 1.40\,$Z_{\odot}$ relation as IRX$=1.44\times [10^{0.4(2.14\beta+5.10)\mu}$].
		 The \citet{overzier11} curve for Lyman break analogs (LBAs) is also shown for reference. 
		 Our low-metallicity stacks agree well with the SMC, 0.14\,$Z_{\odot}$ curve, while the high-metallicity stacks lie closer to the C00, 1.40\,$Z_{\odot}$ and the MHC99 curves.
		}
		\label{fig:environ}
\end{figure*}	

\section{Sample and Data} \label{sec:sample}
Our sample is part of the spectroscopic survey of protocluster and field galaxies in the COSMOS and UDS fields presented in \citet{darvish20}. The primary sample is selected from the narrow-band {\halpha} emitting catalog of High-$z$ Emission Line Survey \citep[HiZELS;][]{sobral14} with followup spectroscopy in $K$-band using Keck/MOSFIRE (PI: N. Scoville). Filler targets are selected to be star-forming galaxies (using color-color selections, e.g., NUV-r-J; \citealt{ilbert13}) with photometric redshifts at $z_{\rm phot}\sim 1.7-2.8$. The parent sample consists of 30 protocluster members at $z\sim 2.2$ and 217 field galaxies.
For survey details and protocluster identification refer to \citet{darvish20} and Sattari et~al. (in preparation). 
We use the spectra and photometry to extract key parameters, as explained below, with uncertainties estimated as the standard deviation of 1000 Monte-Carlo realizations, unless otherwise stated.

\paragraph{Line measurements and metallicities}
The 1D spectra are fitted with a triple Gaussian function for the {\halpha} and the [N{\sc ii}]$\lambda 6550, 6585$ lines (Sattari et~al., in preparation).
Oxygen abundances ($12+\log({\rm O/H})$), or metallicities, are calculated from the ratio of [N{\sc ii}]$\lambda 6585$ to {\halpha}, assuming the empirical calibrations of \citet{pp04}. The ratio is a good metallicity tracer in the local universe \citep[e.g.,][]{marino13}. However, as the calibrations are uncertain at high redshifts \citep[e.g.,][]{bian18}, the absolute metallicity values reported here should be taken with caution. We emphasize that using the [N{\sc ii}]-to-{\halpha} ratio is robust when dividing galaxies into broad metallicity categories discussed in this paper.

\paragraph{Photometry and SED parameters} Catalogs of COSMOS2015 \citep{laigle15} and SPLASH-SXDF \citep[v1.5;][]{mehta18} are used for sources in COSMOS and UDS, respectively, to calculate UV properties (see below), and infer SED parameters. 
For the SED fitting, both catalogs use the \citet{bc03} templates, assuming a Chabrier IMF, a range of stellar metallicities, exponentially declining SFHs (the COSMOS15 catalog also assume delayed SFHs), and two attenuation curves, the C00 curve and a steeper curve (similar to the SMC curve). For details, we refer to the survey papers. 

\paragraph{UV luminosity and $\beta$} To derive the UV continuum slope, $\beta$, and UV luminosity density ($L_{\nu}$) at 1600\,{\AA}, we fit a power-law to the photometry at rest-frame $\lambda=1260-2600$\,{\AA}, $f_{\lambda}\propto \lambda^{\beta}$ (COSMOS and UDS have good rest-UV coverage with at least four photometric points in this range). We avoid the wavelength range of $\lambda=1950-2400$\,{\AA}, as it may potentially be affected by the 2175\,{\AA} UV bump, seen in the attenuation curve of high-metallicity galaxies at these redshifts \citep{scoville15,battisti20,shivaei20}. 
Hereafter, $\nu L_{\nu}$ at 1600\,{\AA} is referred to as UV luminosity, L(UV).

\begin{figure*}[ht]
	\centering
		\includegraphics[width=.9\textwidth,trim={.2cm 0 0 0},clip]{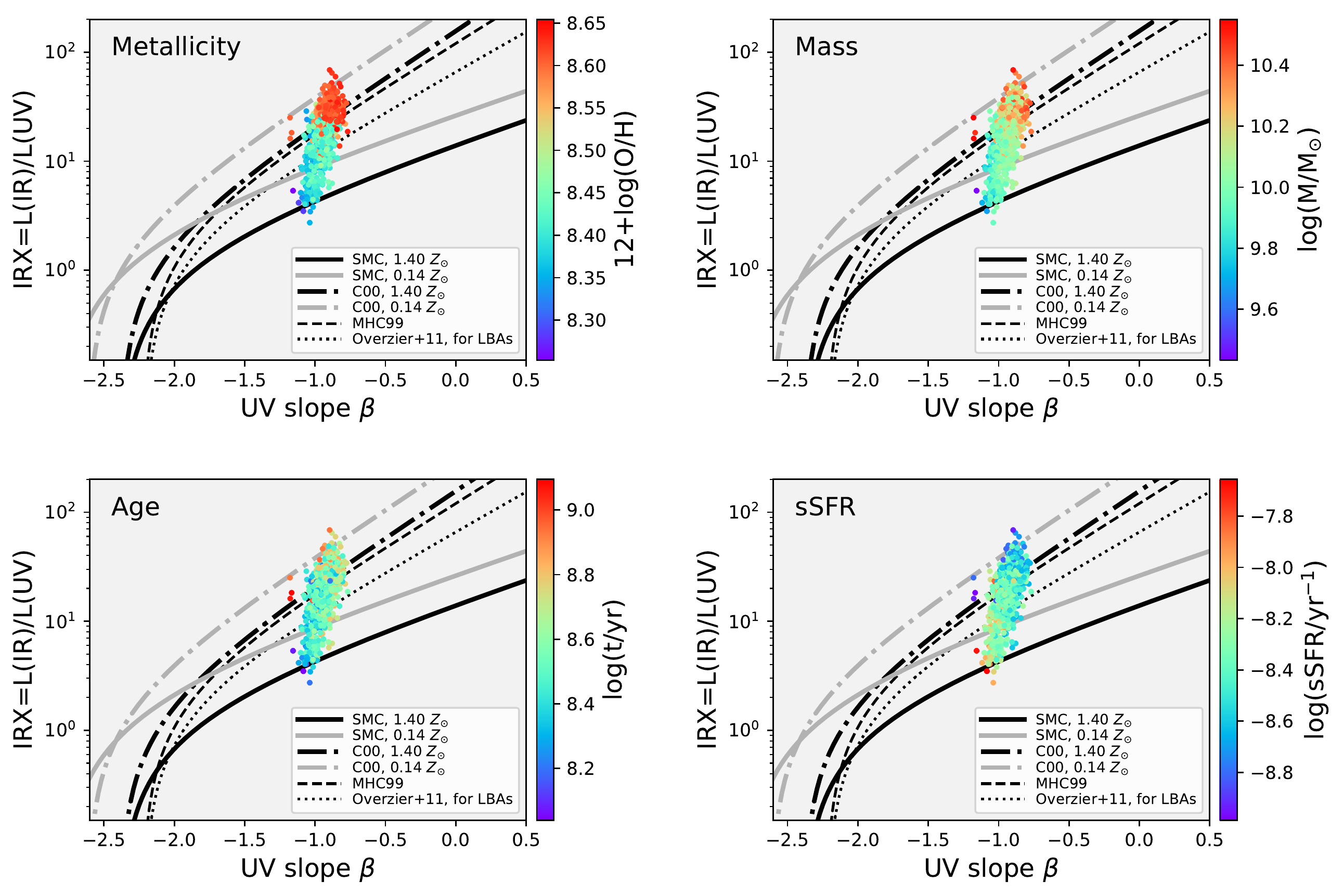} 
		\caption{IRX-$\beta$ at $\beta\sim [-1.2,-0.7]$ color-coded by metallicity, mass, age, and sSFR, as indicated. Points are measured for bootstrapped samples in different bins of metallicity, as explained in Appendix~\ref{ap:a3}.
		Curves are the same as in Figure~\ref{fig:environ}. 
		}
		\label{fig:irx-beta-color}
	\vspace{20pt}
\end{figure*}

\paragraph{IR luminosity} We use the public Spitzer/MIPS 24\,{\um} images of the COSMOS Spitzer survey \citep[S-COSMOS;][]{sanders07}\footnote{\url{https://irsa.ipac.caltech.edu/data/COSMOS/gator_docs/scosmos_mips_24_go3_colDescriptions.html}} and the Spitzer UKIDSS Ultra Deep Survey (SpUDS; PI: J. Dunlop)\footnote{\url{https://irsa.ipac.caltech.edu/data/SPITZER/SpUDS/images}}. For Herschel/PACS data, we use the public images of the PACS Evolutionary Probe survey \citep[PEP;][]{lutz11}\footnote{\url{https://irsa.ipac.caltech.edu/data/COSMOS/overview.html}} in COSMOS and those of the fourth Hershel-SPIRE/SAG-1/Herschel Multi-tiered Extragalactic Survey \citep[HerMES;][]{oliver12}\footnote{\url{https://hedam.lam.fr/HerMES/index/dr4}} in UDS. The pixel scale of MIPS 24\,{\um} images is $1.^{\prime\prime}2\times 1.^{\prime\prime}2$. We matched the pixel scales of PEP PACS images to those of HerMES PACS images, by smoothing them to $2.^{\prime\prime}4\times 2.^{\prime\prime}4$ and $3^{\prime\prime}\times 3^{\prime\prime}$ for the 100\,{\um} and 160\,{\um} images, respectively.

Due to the confusion and sensitivity limits of MIPS and PACS data, stacking is required to detect the IR emission of typical galaxies at $z\gtrsim 1$. Therefore, we stack the target images and perform aperture photometry, as described in Appendix~\ref{ap:a1}.
To estimate the total IR luminosity, we fit the photometry with two sets of IR templates: (i) the locally-calibrated IR templates of \citet[][hereafter R09]{rieke09} and, (ii) a library of the local low-metallicity galaxy templates from \citet{lyu16}. The R09 LIRGs templates accurately fit our galaxies with high metallicities ($12+\log({\rm O/H})\gtrsim 8.5$), in agreement with other studies that support the use of local LIRG templates for massive, IR-bright galaxies at high redshifts \citep[e.g.,][]{derossi18}. However, in Appendix~\ref{ap:a2} we show that the lower metallicity galaxies fit better with the low-metallicity local templates.
Ultimately, the best-fit template is determined through a minimum $\chi^2$ method and the total IR luminosity is calculated by integrating the best-fit template at $\lambda= 8-1000$\,{\um}. 

\paragraph{Final sample} The final sample consists of objects with $> 3\sigma$ detection in {\halpha} and [N{\sc ii}], required for robust metallicity calculations. We remove targets that are classified as mergers by visual inspection of the spectra and/or images, IR AGN (using the \citealt{donley12} criteria), X-ray AGN (from the COSMOS and UDS catalogs), and optical AGN ([N{\sc ii}]/{\halpha}$> 0.5$). Furthermore, as ULIRGs are known to be outliers in the IRX-$\beta$ relation \citep[e.g.,][]{casey14}, we remove targets with $> 3\sigma$ detection in Herschel/PACS 100\,{\um} or 160\,{\um} (corresponding to galaxies with $L({\rm IR})\gtrsim 10^{12}\,L_{\odot}$). 
The final sample consists of 62 field and 13 protocluster galaxies at $z=2.0-2.5$.

\begin{figure*}[ht]
	\centering
		\includegraphics[width=.9\textwidth,trim={.2cm 0 0 0},clip]{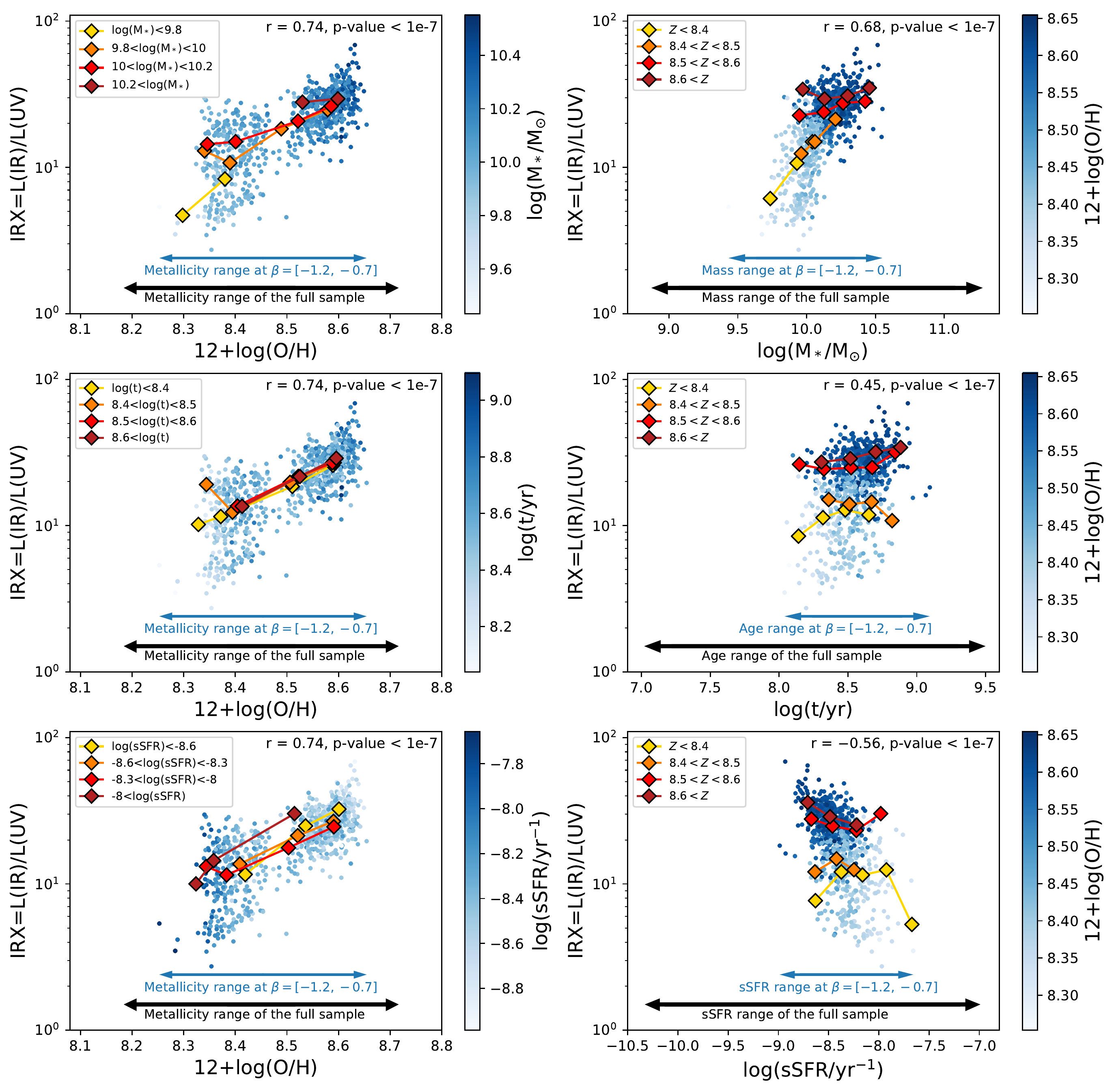} 
		\caption{
		Left column: IRX versus metallicity at $-1.2<\beta<-0.7$ for the measurements in Figure~\ref{fig:irx-beta-color} (blue circles). The average values in narrow bins of mass, age, and sSFR (top to botton, respectively) are shown with diamonds. The Pearson correlation factors (r) and corresponding p-values for the blue points are also shown. The ranges of the parameters at $-1.2<\beta<-0.7$ and those of the full sample are shown with blue and black arrows, respectively. Right column from top to bottom: IRX versus mass, age, and sSFR with averages in narrow bins of metallicity. Symbols are the same as the left column.}
		\label{fig:parcorr}
	\vspace{20pt}
\end{figure*}

\section{IRX-$\beta$ as a function of environment and galaxy properties } \label{sec:res}

Average IRX values are derived in two metallicity bins divided at {\oh}\,$=8.5$, and four $\beta$ bins at $\langle\beta\rangle\sim -1.4, -0.9, -0.5, 0.2$. We choose the metallicity limit of {\oh}\,$=8.5$ as it has been shown that $z\sim 2$ galaxies with metallicities below and above this limit have different attenuation curves \citep{shivaei20}. The exact $\beta$ boundaries are chosen randomly from the ranges shown in Figure~\ref{fig:environ}-left for 100 trials. The details of sampling are described in Appendix~\ref{ap:a3}. We later exclude the reddest $\beta$ bin that includes galaxies with $\beta\gtrsim 0$ from the analysis, as their $\beta$s are uncertain and likely affected significantly by the older stellar populations, owing to their lower sSFRs (average of 0.09\,${\rm Gyr}^{-1}$) and older ages (average of 1\,Gyr) compared to the rest of the sample.

Figure~\ref{fig:environ}-right shows the IRX-$\beta$ in the two bins of metallicity for the full sample (field and protocluster galaxies), and for the field galaxies only. We do not find any significant difference between the IRX-$\beta$ values of the full sample and those of the field galaxies only, suggesting that the trend is independent of the environment. On the other hand, the low-metallicity galaxies in both samples agree better with the SMC relations, while the high-metallicity ones at the same $\beta$ favor the C00 and MHC99 relations. Figure~\ref{fig:environ} shows the range of IRX-$\beta$ relations for different attenuation curves (with slopes in between the SMC and C00 slopes), and different intrinsic $\beta_0$ values. 

The change in IRX with metallicity is less significant at $\beta\lesssim -1.2$, both because the IRX-$\beta$ curves are less distinguishable in this range, and also due to our sample incompleteness at these blue $\beta$s. The HiZELS parent sample at $z\sim 2$ is complete down to $M_*\sim 10^{9.7}$\,{\msun} \citep{sobral14}, which corresponds to $\beta\sim -1.3$ \citep{fudamoto19} and {\oh}\,$\sim 8.3$ \citep{sanders18}. As we specifically select emission-line detected galaxies, the average IRX at $\beta<-1.3$ and {\oh}\,$<8.3$ is possibly biased towards higher values, suffering from a Malmquist bias. 

We further investigate the IRX-$\beta$ scatter as a function of other galaxy properties at $\beta\sim -1.2$ to $-0.7$, as at this $\beta$ range our sample is complete and includes galaxies with a wide range of metallicities.
We divide the sample into multiple metallicity bins, and bootstrap resample from each bin to stack the IR data and calculate IRX. We refer to Appendix~\ref{ap:a3} for the sampling details.
Panels of Figure~\ref{fig:irx-beta-color} show the IRX-$\beta$ measurements color-coded with metallicity, mass, age, and sSFR.
IRX increases significantly with increasing metallicity at a given $\beta$, such that as metallicity increases, galaxies occupy the full range of attenuation curves from the SMC to the MHC99 and C00 relations. The same trend, but weaker, is seen with increasing stellar mass and age and decreasing sSFR\footnote{The SED-inferred SFRs and ages are highly sensitive to the assumed attenuation curve \citep{shivaei18,shivaei20}. However, our SFRs and ages are less affected, as a range of attenuation curves is assumed in the SED fittings used in this work (Section~\ref{sec:sample}).}.

\section{Which parameter drives the IRX-$\beta$ scatter? }
It is expected, from a physical point of view, that dust attenuation properties correlate with gas abundances. To confirm this statement, we perform a partial correlation test to determine whether the IRX dependence on metallicity remains when the other parameters are held fixed. We divide the realizations in Figure~\ref{fig:irx-beta-color} into a series of bins with narrow ranges in metallicity, mass, age, and sSFR. We calculate the mean IRX in bins of mass, age, and sSFR as a function of metallicity (diamonds in the left column of Figure~\ref{fig:parcorr}), and in bins of metallicity as a function of the other three parameters (diamonds in the right column of Figure~\ref{fig:parcorr}). The left column of Figure~\ref{fig:parcorr} shows that the IRX correlation with metallicity remains unchanged even if we look at it in narrow bins of mass, age, and sSFR. On the other hand, the right panels show that any correlations with mass, age, and sSFR are much weaker when observed in bins of metallicity.
This indicates that among these four parameters, metallicity has the strongest correlation with IRX and UV attenuation.

The Pearson correlation factor and the range of parameters covered at the given $\beta$ reinforce the conclusions we have already drawn.
IRX has the largest correlation factor with metallicity (shown in the corners of plots in Figure~\ref{fig:irx-beta-color}), and the widest metallicity range, comparable to the metallicity range of the full sample. Variations in mass, sSFR, and age move galaxies more significantly along the IRX-$\beta$ curves (i.e., changing both IRX and $\beta$), as opposed to  changing IRX at a given $\beta$ -- which explains the small dynamic ranges of these properties at this $\beta$ bin compared to the range of the properties in the full sample (blue and black arrows in Figure~\ref{fig:parcorr}, respectively).

We conclude that a major factor in the scatter of the IRX-$\beta$ relation is metallicity, as metallicity reflects the properties of small dust grains that determine the steepness of the attenuation curve in the UV \citep[e.g.,][]{gordon97,zelko20}.
At low metallicities, the intense UV radiation in the ISM and the reduced gas and dust shielding from the UV radiation, due to lower molecular fraction (ratio of molecular to total gas mass), make the environment favorable to shatter large dust grains into small ones, resulting in a steep rise in the UV attenuation curve \citep{shivaei20}.

\section{Implications for UV dust corrections at high redshifts} \label{sec:disc}

\begin{figure}[t]
	\centering
		\includegraphics[width=.45\textwidth,trim={.2cm 0 0 0},clip]{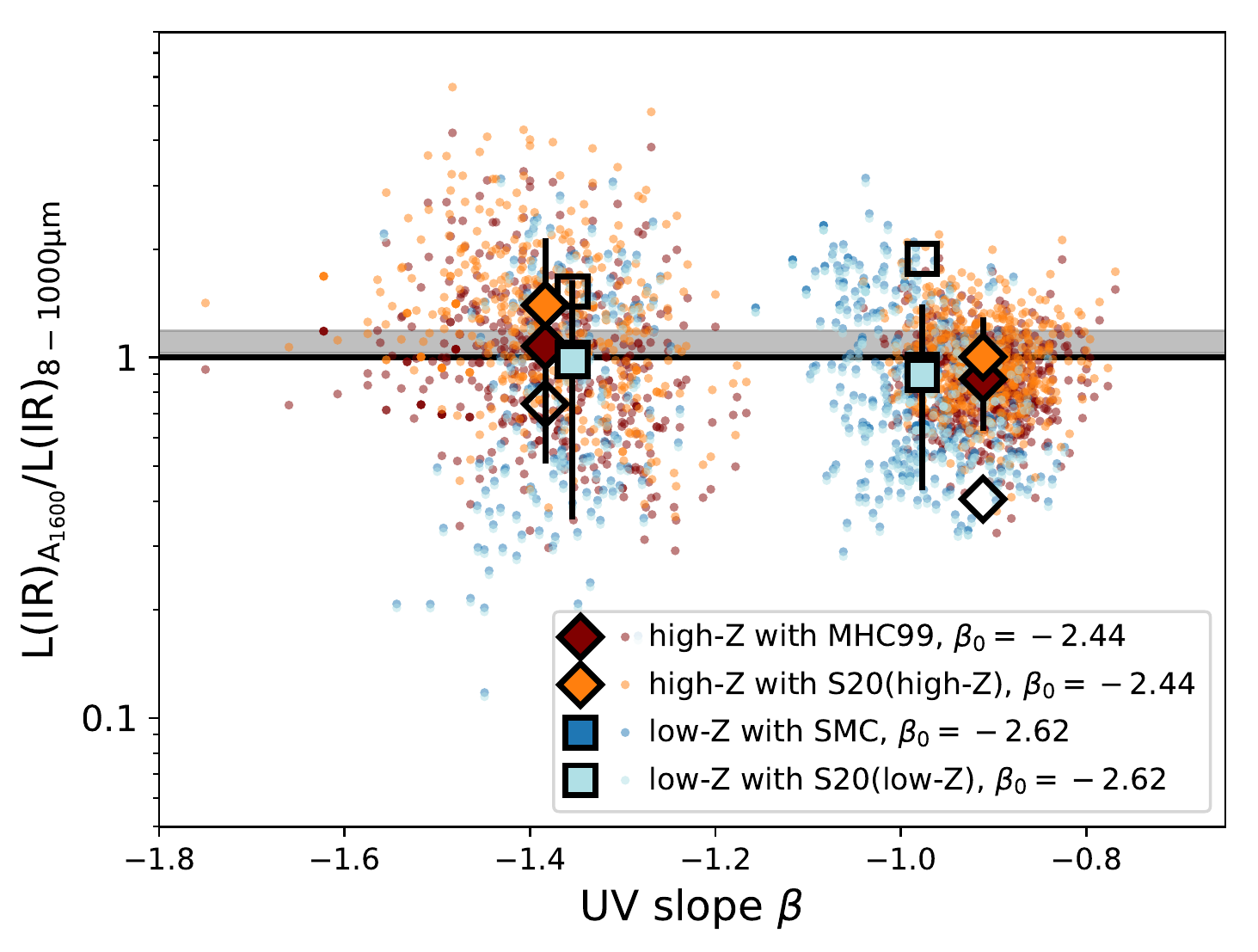}
		\caption{Ratio of predicted ($L({\rm IR})_{\rm A_{1600}}$, Equation~\ref{eq:lir}) to  measured ($L({\rm IR})_{\rm 8-1000{\mu m}}$) IR luminosities as a function of $\beta$. Each small circle is a random realization in metallicity and $\beta$ bins, calculated in the same way as in Figure~\ref{fig:irx-beta-color} with the difference of having a single metallicity division at $12+\log({\rm O/H})=8.5$ (Appendix~\ref{ap:a3}). Averages in each metallicity and $\beta$ bin are shown with large symbols.
		Dark red and orange symbols show high-metallicity stacks, for which $L({\rm IR})_{\rm A_{1600}}$ is calculated using the MHC99 relation and the high-metallicity relation of \citet[][S20]{shivaei20}, respectively. Dark blue and light blue symbols (which are indistinguishable) show low-metallicity stacks, for which $L({\rm IR})_{\rm A_{1600}}$ is calculated using the SMC relation from \citet{reddy18a} and \citet{gordon03} and the S20 low-metallicity relation, respectively. 
		The horizontal line is unity and the grey region reflects the range of different bolometric corrections ($\gamma\sim 1.79-2.07$, Equation~\ref{eq:lir}) from \citet{reddy18a} and MHC99.
		The empty squares (diamonds) show the calculations using the MHC99 (SMC) relation for the low-(high-)metallicity galaxies, which result in overestimation (underestimation) of $L({\rm IR})_{\rm A_{1600}}$.
		}
	\vspace{20pt}
		\label{fig:irratio}
\end{figure}

The IRX-$\beta$ relation is a powerful tool for correcting the observed UV luminosities for attenuation. 
Figure~\ref{fig:irratio} shows the ratio of predicted L(IR) from UV observations, to the observed L(IR), integrated at $\lambda=8-1000$\,{\um}. The predicted L(IR), $L({\rm IR})_{\rm A_{1600}}$, is calculated as:
\begin{equation}\label{eq:lir}
L({\rm IR})_{\rm A_{1600}} = \gamma\times L({\rm UV})\times (10^{0.4\,{\rm A}_{1600}} - 1),
\end{equation}
where $L({\rm UV})$ is the observed UV luminosity at 1600\,{\AA}, A$_{1600}$ is the total attenuation at 1600\,{\AA} calculated using $\beta$ and assuming an attenuation curve, and $\gamma$ is the bolometric correction to convert the absorption at 1600\,{\AA} to total absorbed UV flux. We adopt $\gamma = 2.13$ from the IRX calibrations of \citet{hao11} with Balmer decrement-corrected {\halpha} luminosities.
We assume different $\beta-A_{1600}$ relations for galaxies below and above {\oh}\,$=8.5$: two C00-like relations with solar-metallicity populations for the high-metallicity points (MHC99 and the S20 high-metallicity curve), and two relations that follow a steeper SMC-like curve and sub-solar stellar populations (the SMC relation of \citealt{reddy18a} and the S20 low-metallicity curve).  
The good agreement between the predicted and observed L(IR) in Figure~\ref{fig:irratio} indicates that, {\em on average}, galaxies with {\oh} below and above 8.5 (corresponding to $M_*\sim 10^{10.4}$\,{\msun}, owing to the mass-metallicity relation; e.g., \citealt{sanders18}) follow a steep SMC-like and a shallower C00-like curves, respectively. 
In reality, there is an infinite range of attenuation curve possibilities and, based on the results of this study, their slopes correlate with galaxies metallicities. 
We show in Figure~\ref{fig:irratio} (empty symbols) that using a C00-like attenuation curve for the low-metallicity galaxies would overestimate the predicted $L({\rm IR})_{\rm A_{1600}}$ up to a factor of $\sim 3$. 
These results are specifically important for high-redshift studies, as the galaxies at higher redshifts have lower metallicities at a given mass \citep{maiolino19}, and presumably more SMC-like stellar dust attenuation properties.

~\\
\paragraph{\bf Acknowledgment}
IS thanks Robin Ciardullo and Joel Leja for helpful conversations.
IS is supported by NASA through the NASA Hubble Fellowship grant \# HST-HF2-51420, awarded by the Space Telescope Science Institute, which is operated by the Association of Universities for Research in Astronomy, Inc., for NASA, under contract NAS5-26555. 

\bibliographystyle{apj}

\appendix 
\section{A.~~MIPS and PACS Stacking and Aperture Photometry} \label{ap:a1}

\paragraph{Stacking} Due to the confusion and sensitivity limits of Spitzer/MIPS and Herschel, stacking is required to detect the IR emission of typical galaxies at $z\gtrsim 1$. To perform stacking, we construct 40$\times$40 pixel subimages centered at our targets' optical coordinates. When necessary, the images are shifted by sub pixel values to accurately center the targets.
For MIPS images, we use a list of prior sources with signal-to-noise of $>3$ in Spitzer/IRAC 1 and 2. For PACS images, we use a list of priors with signal-to-noise of $>3$ in MIPS 24\,{\um}. Using the prior lists, we model the emission of the companion sources (all of the prior objects, except for the target) by performing scaled point-spread function (PSF) photometry on the subimages. By subtracting the companion model images from the original image, we make clean subimages that only include our target. Using the clean subimages, we perform 3$\sigma$-trimmed (clipped) mean stacking. Trimmed mean is used to ensure the mean is not biased towards outliers.

\paragraph{Aperture photometry} To measure fluxes and their associated errors, we perform aperture photometry on the stacked images. We use an aperture size of $3^{\prime\prime}$, $7.2^{\prime\prime}$, and $9^{\prime\prime}$ for the 24\,{\um}, 100\,{\um}, and 160\,{\um} images, respectively. The best aperture sizes are determined by performing aperture photometry with different aperture sized on the stacked images of all the sources in each band and choosing the aperture with the highest signal-to-noise in each band. For the 24\,{\um} stacks, we measure the aperture correction (the amount of light that is lost outside of the aperture) from the growth curve of the 24\,{\um} stacked image of all the source. For the PACS aperture corrections, we use the calculated PACS Encircled Energy Fractions (EEFs) of \citet[][Table 2]{balog14}. The flux errors are estimated as the standard deviation of the fluxes measured in 100 apertures with the same radii as the source aperture radius, located at random positions away from the center of the image (where the source is) by more than 1\,FWHM of the image PSF.

\section{B.~~IR templates} \label{ap:a2}

\begin{figure*}[t]
	\centering
		\includegraphics[width=.9\textwidth,trim={.2cm 0 0 0},clip]{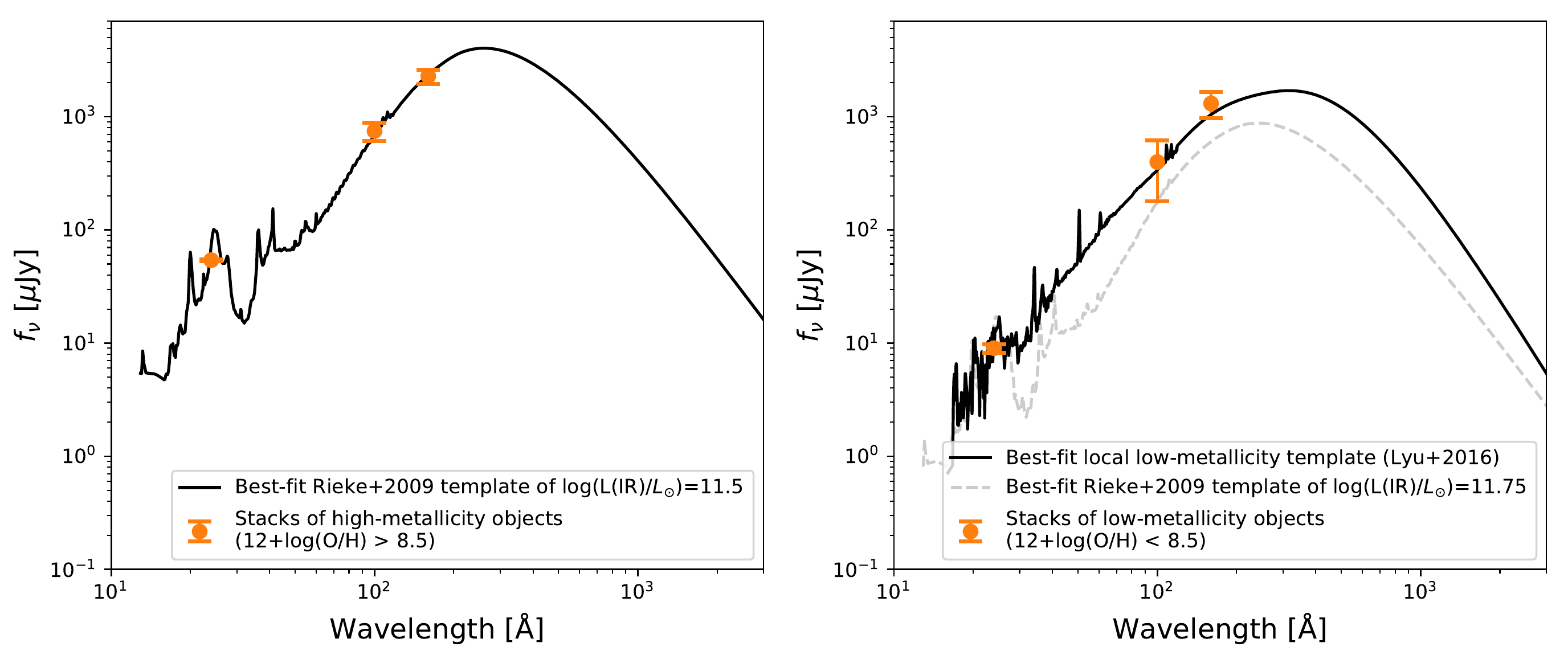} 
		\caption{The best-fit IR SEDs IR to the 24, 100, and 160\,{\um} stacked photometry (orange circles) of our galaxies with metallicities above (left) and below (right) of $12+\log({\rm O/H})=8.5$. The LIRG templates of \citet{rieke09} accurately reproduce the observed photometry of our high-metallicity galaxies (left), while their 7.7\,{\um} PAH intensity is overestimated for the low-metallicity galaxies (gray curve in the right plot). We use a library of local low-metallicity templates from \citet{lyu16} to fit our low-metallicity galaxies (black curve in the right plot).
		}
		\label{fig:irtemplates}
\end{figure*}

We use the locally-calibrated IR templates of \citet[][R09]{rieke09} and a library of the local low-metallicity galaxies template from \citet{lyu16} to fit our high- and low-metallicity galaxies, respectively. The R09 templates are limited to those with luminosities of $10^{10.5}$ to $10^{12}\,L_{\odot}$, as previous studies have shown that these templates accurately match the IR emission of LIRGs at high redshifts\citep[e.g.,][]{derossi18}.\footnote{IR luminosities estimated from other commonly used templates, such as the \citet{ce01} and \citet{elbaz11} templates, are consistent with those derived from the R09 templates (see \citealt{reddy12a} for templates comparison).}
The left panel of Figure~\ref{fig:irtemplates} show the fit to the stacked photometry of our sample with $12+\log({\rm O/H})> 8.5$.

On the other hand, the local LIRG and ULIRG templates are not appropriate for low-metallicity and low-mass galaxies at high redshifts.
\citet{shivaei17} showed that the intensity of the 7.7\,{\um} aromatic band emission, which is traced by MIPS 24\,{\um} at $z\sim 2$, is significantly suppressed at metallicities of $12+\log({\rm O/H})\lesssim 8.5$\footnote{The metallicity here is estimated using the ratio of [N{\sc ii}]/{\halpha} and the calibrations of \citet{pp04}.} compared to that in the higher metallicity galaxies at the same redshift and to that in the locally calibrated templates of LIRGs and ULIRGs. The right panel of Figure~\ref{fig:irtemplates} shows that the R09 templates can not reproduce the observed 24, 100, and 160\,{\um} photometry of sources with $12+\log({\rm O/H})< 8.5$. Therefore, for our low-metallicity galaxies, we use a library of 19 templates of local low-metallicity galaxies from \citet{lyu16}. These galaxies were selected from the Dwarf Galaxy Survey \citep[DGS;][]{madden13} to have high enough signal-to-noise Spitzer/IRS spectra. The IR templates are constructed by combining Herschel and WISE photometry with Spitzer/IRS spectra. The right panel of Figure~\ref{fig:irtemplates} shows how well these templates fit our observed low-metallicity photometry. If the 24\,{\um} data is excluded from the fits, the inferred IR luminosities from the best-fit local low-metallicity templates and those from the best-fit R09 templates are consistent with each other within the uncertainties.

\section{C.~~Sampling Methodology} \label{ap:a3}

To avoid introducing biases in the results by arbitrarily choosing fixed $\beta$ and metallicity boundaries, we resample the data in many trials using different metallicity and $\beta$ boundaries, as follows. In Figure~\ref{fig:environ}, the sample is divided into two metallicity bins at {\oh}\,$=8.5$ and into four $\beta$ bins (however the reddest $\beta$ bin is later excluded from the analysis).
We perform 100 trials to randomly select boundaries of $\beta$ from the following intervals: $\beta=[-1.25,-1.05]$, $[-0.8,-0.6]$, and $[-0.2,0.1]$ (Figure~\ref{fig:environ}-left). The width of the $\beta$ ranges are chosen to reflect the typical uncertainty of $\beta$ in each range. Then for each realization, we stack the IR images, choose the IR template with the least $\chi^2$ value from the best-fit R09 template and the best-fit low-metallicity template, and calculate the IR luminosity. If all three of the MIPS and PACS stacks are undetected for a subsample, that realization is removed from the analysis due to unconstrained IR luminosity. The IRX value for each realization is the ratio of the IR luminosity to $3\sigma$-trimmed average UV luminosity. 
Finally, the average and standard deviation of the IRX distribution are taken as the IRX and its error, respectively, and the average $\beta$ and metallicity are calculated in Figure~\ref{fig:environ}-right.

The same sampling procedure is used for the datapoints in Figures~\ref{fig:irx-beta-color} and \ref{fig:parcorr}, with two differences: 1) to avoid biases due to arbitrarily chosen metallicity bins, we use five different metallicity boundaries at {\oh}\,$=8.35$ to 8.55 with 0.05\,dex intervals, and 2) we bootstrap resample from each bin to stack the IR data and calculate IRX. The exact minimum and maximum $\beta$s in each metallicity bin are randomly chosen from the ranges of $\beta=[-1.25,-1.05]$ and $[-0.8,-0.6]$, respectively, for each bootstrapped sample (repeated 100 times). Then, IRX and average $\beta$, metallicity, mass, age, and specific SFR (sSFR$=$SFR/M$_*$) are calculated for each realization, and shown as circles in Figures~\ref{fig:irx-beta-color} and \ref{fig:parcorr}. Figure~\ref{fig:irratio} shows the same bootstrapped realizations but divided at a single metallicity boundary and into two $\beta$ bins.

\end{document}